\documentclass[sigconf,nonacm]{acmart}
\AtBeginDocument{%
  }

\usepackage[T1]{fontenc}
\usepackage{graphicx}
\usepackage[show]{chato-notes}
\usepackage{booktabs}
\usepackage{xspace}
\usepackage{pifont}
\usepackage{amsmath} 
\usepackage{booktabs}
\usepackage{multirow}
\usepackage{siunitx}
\usepackage{subcaption}
\usepackage{threeparttable}
\usepackage{hyperref}

\usepackage{amssymb}
\usepackage{makecell}
\usepackage{mathtools}
\usepackage{enumitem}
\usepackage{adjustbox}
\usepackage{booktabs} 
\usepackage{xcolor}   
\usepackage{cancel}
\newcommand{\msmarcovo}{\textsc{Ms Marco-v1}\xspace}

\newcommand{\lotte}{\textsc{LoTTE}\xspace}

\newcommand{\kannolo}{\textsc{kANNolo}\xspace}
\newcommand{\faiss}{\textsc{Faiss}\xspace}
\newcommand{\warp}{\textsc{Warp}\xspace}
\newcommand{\emvb}{\textsc{Emvb}\xspace}
\newcommand{\igp}{\textsc{Igp}\xspace}

\newcommand{\plaid}{\textsc{Plaid}\xspace}

\newcommand{\hnsw}{\textsc{Hnsw}\xspace}
\newcommand{\fkmeans}{\textsc{FastKMeans}\xspace}
\newcommand{\fkmeansrs}{\textsc{FastKMeans-rs}\xspace}

\newcommand{\splade}{\textsc{Splade}\xspace}

\newcommand{\colbert}{\textsc{ColBERT}\xspace}
\newcommand{\colbertvt}{\textsc{ColBERTv2}\xspace}
\newcommand{\xtr}{\textsc{Xtr}\xspace}

\newcommand{\cocond}{\textsc{CoCondenser}\xspace}

\newcommand{\tac}{\textsc{Tac}\xspace}

\newcommand{\tachiom}{\textsc{Tachiom}\xspace}

\begin{document}

\title{Efficient Multivector Retrieval with Token-Aware Clustering and Hierarchical Indexing}


\author{Silvio Martinico}
\authornote{All authors contributed equally to this research.}
\orcid{0009-0005-7280-6147}
\affiliation{%
  \institution{University of Pisa \& ISTI--CNR}
  \city{Pisa}
  \country{Italy}
}
\email{silvio.martinico@phd.unipi.it}

\author{Franco Maria Nardini}
\orcid{0000-0003-3183-334X}
\affiliation{%
  \institution{ISTI--CNR}
  \city{Pisa}
  \country{Italy}
}
\email{francomaria.nardini@isti.cnr.it}

\author{Cosimo Rulli}
\orcid{0000-0003-0194-361X}
\affiliation{%
  \institution{ISTI--CNR}
  \city{Pisa}
  \country{Italy}
}
\email{cosimo.rulli@isti.cnr.it}

\author{Rossano Venturini}
\orcid{0000-0002-9830-3936}
\orcid{1234-5678-9012}
\affiliation{%
  \institution{University of Pisa}
  \city{Pisa}
  \country{Italy}
}
\email{rossano.venturini@unipi.it}

\renewcommand{\shortauthors}{Silvio Martinico, Franco Maria Nardini, Cosimo Rulli, and Rossano Venturini}

\begin{abstract}
Multivector retrieval models achieve state-of-the-art effectiveness through fine-grained token-level representations, but their deployment incurs substantial computational and memory costs. Current solutions---based on the well-known $\kappa$-means clustering algorithm---group similar vectors together to enable both effective compression and efficient retrieval. However, standard $\kappa$-means scales poorly with the number of clusters and dataset size, and favours frequent tokens during training while underrepresenting rare, discriminative ones. In this work, we introduce \tachiom, a multivector retrieval system that exploits token-level structure to significantly accelerate both clustering and retrieval. By accounting for tokens' distribution during centroid allocation, \tachiom\ easily scales to millions of centroids, enabling highly accurate document scoring using only centroids, avoiding expensive token-level computation. \tachiom\ combines a graph-based index over centroids with an optimized Product Quantization layout for efficient final scoring. Experiments on \msmarcovo\ and \lotte\ show that \tachiom\ achieves up to $247\times$ faster clustering than $\kappa$-means and up to $9.8\times$ retrieval speedup over state-of-the-art systems while maintaining comparable or superior effectiveness.
\end{abstract}




\maketitle

\section{Introduction}
\label{sec:intro}

Transformer-based language models~\cite{bert,attention,dpr,formal2021splade,lassance2024splade} have fundamentally reshaped Information Retrieval (IR), shifting the paradigm from lexical-based to representation-based retrieval. Within this landscape, multivector models, such as ColBERT~\cite{colbert,colbertv2}, have emerged as a gold standard for effectiveness. By encoding documents as sets of token-level vectors and computing relevance via late interaction, i.e., allowing every query token to interact with every document token, these models capture fine-grained semantic nuances that single-vector dense retrievers often miss.
However, the real-world deployment of multivector models remains severely constrained by their computational demands. Storing and searching large sets of token vectors creates massive memory footprints and computational overhead. To mitigate this, state-of-the-art solutions adopt a \textit{gather-and-refine} strategy combined with token quantization~\cite{dessert,plaid,emvb,warp,igp}. In the gather phase, a lightweight search over token vectors identifies promising candidates; in the refine phase, full MaxSim scores are computed only for the selected candidates. Both phases critically depend on a coarse quantizer---a set of centroid vectors that approximate the token embedding space. During gathering, centroids serve as efficient proxies for locating relevant tokens. For compression, each token vector is decomposed into a centroid assignment and a residual vector, which is then compressed using more precise Vector Quantization (VQ) techniques, like, for instance, Product Quantization~\cite{pq} (PQ). This two-level scheme substantially improves approximation: centroid assignments distinguish distant vectors at minimal cost, while residual compression captures local nuances.

\colbertvt~\cite{colbertv2} pioneered this centroid-based approach, compressing the residual (elementwise difference between the vector and the centroid) rather than the vector itself. \plaid~\cite{plaid} extended this framework with centroid interaction and inverted-file filtering over clustered token embeddings for efficient retrieval.
Subsequent work refined this framework: \emvb~\cite{emvb} introduced bitvector prefiltering to accelerate candidate selection and optimized PQ variants~\cite{opq,jmpq} to improve redisuals' compression, achieving significant speedups over \plaid. \warp~\cite{warp} combined \plaid's centroid interaction with \xtr's lightweight architecture~\cite{xtr}; the authors showed that their method also generalize to \colbertvt embeddings. \igp~\cite{igp} proposed building a graph index on the set of centroids to avoid exhaustive centroids scoring. 

Across all these methods, both retrieval quality and compression effectiveness scale directly with the number of centroids: finer-grained centroids enable more precise approximations and more selective gathering. However, standard $\kappa$-means clustering---the predominant method for centroids selection---creates a severe bottleneck: it scales poorly with both dataset size and number of clusters, and critically, it ignores the token structure of multivector embeddings. Training even moderately large sets of centroids (e.g., $262$K) on multivector collections can require tens of hours on multi-core CPUs or expensive GPU resources, fundamentally limiting achievable granularity.
In this paper, we address this bottleneck by exploiting a key structural property of multivector embeddings: \textit{token identity}. Unlike generic vector sets, multivector collections exhibit strong correlations between token type and embedding distribution. Common tokens (e.g., stopwords) dominate $\kappa$-means' loss during training, consuming the majority of centroids, while rare domain-specific tokens---often more discriminative for retrieval---receive minimal representation. We introduce \tachiom, a multivector retrieval system that exploits this structure at both the clustering and retrieval stages. At its core, \tachiom employs Token-Aware Clustering (\tac), which explicitly allocates centroids based on token frequency and semantic variance. By partitioning the clustering workload across token groups, \tac achieves significantly faster centroids computation over standard $\kappa$-means. This efficiency enables scaling to millions of centroids, far beyond prior work. Leveraging these high-resolution centroids, \tachiom performs gathering using only centroid-level interactions via an \hnsw graph index~\cite{hnsw}, so as to bypass expensive token-level computations entirely. For refinement, a cache-optimized PQ layout tailored for late interaction enables efficient final scoring.

\vspace{1mm}
In summary, the novel contributions of this work are:

\begin{itemize}[leftmargin=*]
\item We propose \tac, a clustering algorithm that leverages token identity to decompose the global clustering problem into independent per-token subproblems, improving scalability while producing centroids better aligned with retrieval objectives. We further derive a theoretical lower bound on the computational speed-up achieved by \tac\ over $\kappa$-means.
\vspace{1mm}
\item We introduce \tachiom, a retrieval architecture that leverages the clustering granularity enabled by \tac to perform centroids-only gathering via graph-based search and efficient full scoring with late interaction-tailored PQ.
\vspace{1mm}
\item We conduct comprehensive experiments on \msmarcovo\ and \lotte, demonstrating that our proposed architectures offer superior efficiency-effectiveness trade-offs. Our results show that \tac\ achieves up to $247\times$ faster training than standard $\kappa$-means, while \tachiom\ delivers up to $9.8\times$ faster end-to-end retrieval compared to state-of-the-art systems. To favor reproducibility, we release our implementation of \tac and \tachiom in Rust on GitHub at \href{https://github.com/TusKANNy/tachiom}{\texttt{https://github.com/TusKANNy/tachiom}}.
\end{itemize}

\section{Methodology}
\label{sec:meth}


\subsection{Token-Aware Clustering}
\label{sec:tac}
Multivector models such as \colbert~\cite{colbert} represent each document as a set of contextualized token embeddings. State-of-the-art retrieval methods~\cite{plaid,colbertv2,emvb,warp,igp,jmpq} rely on centroids to approximate these large collections of token vectors, typically using $\kappa$-means clustering to select representative centroids. However, standard $\kappa$-means treats token embeddings as a generic vector set, discarding the structural information provided by token identity.
Token-Aware Clustering (\tac) explicitly exploits the token structure of multivector representations. \tac serves two purposes: (i) it significantly accelerates clustering by partitioning the workload into independent per-token subproblems, crucial for the massive vector collections produced by \colbert-like models; (ii) it balances centroid allocation toward rare tokens, which are underrepresented by standard $\kappa$-means despite their importance for retrieval accuracy.

$\kappa$-means aims to find a set of clusters $ \mathcal{C} = \{C_1, \dots, C_{\kappa}\}$ that partitions the dataset and minimizes the Within-Cluster Sum of Squares (WCSS) or Inertia, i.e.: $$\sum_{i=1}^{\kappa} \sum_{\mathbf{t} \in C_i} \|\mathbf{t} - \mathbf{c}_i\|^2 \ ,$$ where $\mathbf{c}_i$ denotes the centroid of cluster $C_i$.
This objective treats all vectors equally, but multivector collections exhibit severe frequency imbalance. Common tokens (e.g., stopwords) may appear millions of times, while domain-specific terms have way lower frequencies (hundreds or few thousands of occurrences). Analysing two \colbertvt multivector collections, we observed that the top 100 most frequent tokens account for over 40\% of all vectors, specifically 41\% on \msmarcovo and 45\% on \lotte-pooled. Consequently, the WCSS loss is dominated by high-frequency tokens, which consume the majority of centroids despite contributing little to retrieval quality. Meanwhile, rare tokens---which often carry the most discriminative semantic signal~\cite{term_freq}---receive minimal representation, degrading approximation quality precisely where it matters most.

Drawing from classic Information Retrieval literature, where \emph{Inverse Document Frequency} (IDF) has long been employed to favor rare, discriminative terms~\cite{idf_robertson,idf2}, we propose \tac to address this imbalance in the clustering context. \tac decouples centroid allocation from uniform WCSS minimization by explicitly weighting tokens based on their frequency and semantic variance, so as to favor rare tokens. \tac first computes a weight for each token type representing its share of the total centroid budget, then performs independent $\kappa$-means clustering for each token type.

\tac distributes the centroid budget through a four-stage pipeline: (1) \emph{Tail Handling} isolates very rare tokens to prevent their complete marginalization; (2) \emph{Damped Scoring} computes an allocation weight for the remaining tokens; (3) \emph{Bounding} enforces strict limits to avoid both under- and over-allocation; and (4) \emph{Budget Reconciliation} adjusts the final assignments to exactly match the global centroid budget.

\vspace{1mm}
\noindent \textbf{Phase 1: Tail Handling}. We partition token types into three categories based on their frequency $n$: \textit{Micro tokens} ($n < \mu$) receive 1 centroid each, \textit{Small tokens} ($\mu \leq n < \tau$) receive 2 centroids each, and \textit{Active tokens} ($n \geq \tau$) are allocated centroids dynamically based on their frequency and semantic variance. This fixed allocation prevents the complete marginalization of extremely rare tokens.

\vspace{1mm}
\noindent \textbf{Phase 2: Damped Scoring}. For active tokens, we compute a \textit{spread measure} $s_j$ that quantifies semantic variance; we adopt component-wise variance for computational efficiency, i.e.,
$$s_j \coloneqq \frac{1}{n_j} \sum_{i=1}^{n_j} \|\mathbf{t}_{j,i} - \bar{\mathbf{t}}_j\|^2 \ , $$
where $n_j$ is the frequency of token $j$, $\mathbf{t}_{j,i}$ is the $i$-th occurrence of token $j$, and $\bar{\mathbf{t}}_j$ is its mean embedding. We then compute a \textit{damped weight}, 
$\text{w}_j = \sqrt{n_j} \cdot s_j$.
The square-root damping applies diminishing returns to highly frequent tokens, preventing them from monopolizing centroids budget; the spread coefficient $s_j$ favors tokens with high semantic variance. The number of centroids $\kappa_j$ of token $j$ is allocated proportionally to its weight,
$$\kappa_j = \left\lfloor \frac{w_j}{\sum_{i=1}^{N_T} w_i} \cdot B \right\rfloor \ ,$$
where $N_T$ is the number of different tokens in our collection and $B$ is the remaining centroid budget after tail handling.

\vspace{1mm}
\noindent \textbf{Phase 3: Bounding}. We enforce a hard floor $\varepsilon$ such that $\kappa_j \geq \varepsilon$ for all active tokens to guarantee minimum representation. Additionally, we impose an upper bound to prevent over-allocation: each token must satisfy $\frac{n_j}{\kappa_j} \geq \theta$, ensuring at least $\theta$ vectors per centroid on average. This dual bounding prevents both under-representation of rare tokens in low-budget regimes and over-allocation to already well-approximated tokens in high-budget regimes.

\vspace{1mm}
\noindent \textbf{Phase 4: Budget Reconciliation}. We redistribute surplus or deficit centroids to exactly match the target budget $\kappa$. Once centroids are allocated, we perform independent $\kappa_j$-means clustering for each token $j$ with its assigned $\kappa_j$ centroids. 

In summary, \tac fundamentally differs from standard $\kappa$-means in two key aspects: 1) it prioritizes retrieval quality over pure reconstruction error: by allocating centroids based on token frequency and semantic variance rather than uniform WCSS contribution, \tac ensures that rare, discriminative tokens receive adequate representation in terms of assigned centroids; 2) instead of optimizing a single global objective over all $N$ vectors with $\kappa$ centroids, \tac\ solves independent subproblems for each token, clustering the $n_j$ vectors of token $j$ into $\kappa_j$ centroids. This decomposition substantially reduces the per-iteration computational cost and enables effective parallelization, as each token can be processed independently, whereas standard $\kappa$-means requires shared access to the full dataset. Moreover, the damped centroid allocation mitigates the bottleneck caused by highly frequent tokens, further improving both total cost and parallel efficiency.

\vspace{1mm}
\noindent \textbf{Theoretical Lower Bound for Speed-up}. 
Standard $\kappa$-means requires $O(I \cdot N \cdot \kappa \cdot d)$ operations for $I$ iterations over $N$ vectors with $\kappa$ centroids in $d$ dimensions. \tac reduces this to $O(I \cdot \sum_{j=1}^{N_T} n_j \cdot \kappa_j \cdot d)$.

We can provide a minimum theoretical speedup of \tac over $\kappa$-means. Ignoring floor operations and tail handling (Phase 1) for clarity, we have a total centroids budget $\kappa$ and the resulting number of centroids assigned to token $j$ becomes $\kappa_j =  \frac{w_j}{\sum_{i=1}^{N_T} w_i} \cdot \kappa$. Thus, the speedup achieved by \tac over $\kappa$-means is:

\begin{align*}
    \frac{\kappa N}{\sum\limits_{j=1}^{N_T} \kappa_j n_j} 
    \ &= \ \frac{\kappa N}{\sum\limits_{j=1}^{N_T} (w_j/\sum\limits_{i=1}^{N_T} w_i) \kappa n_j} 
    \ = \ \frac{\kappa \left( N \sum\limits_{j=1}^{N_T} w_j \right)}{\kappa\left(\sum_{j=1}^{N_T} w_j n_j\right)} 
    \\ \\
    \ &\geq \ \frac{N \sum\limits_{j=1}^{N_T} w_j}{\max\limits_{j=1}^{N_T} w_j \sum\limits_{j=1}^{N_T} n_j} 
    \ = \ \frac{N \left( \, \sum\limits_{j=1}^{N_T} w_j \right)}{\left( \max\limits_{j=1}^{N_T} w_j \right) N} 
    \ = \ \frac{\sum\limits_{j=1}^{N_T} w_j}{\max\limits_{j=1}^{N_T} w_j } \ .
\end{align*}

\smallskip
That is, the speedup is lower-bounded by the ratio of the total weight to the maximum single token weight. Because the cumulative weight across the vocabulary vastly exceeds that of any single token, this formulation guarantees a substantial baseline speedup. Furthermore, this bound reveals a key insight: our damped scoring mechanism not only balances centroid allocation toward rare tokens, but actively suppresses the dominance of the most common ones. By applying square-root scaling to frequencies, \tac directly reduces the maximum single-token weight, strictly improving the speedup guarantee and preventing these dominant tokens from creating a computational bottleneck.

\vspace{1mm}
\noindent \textbf{Residuals Compression}. 
After computing the centroids, we compress the residuals, i.e., the difference between each token vector in the dataset and its assigned centroid, using PQ~\cite{pq}. Because residual magnitudes may vary across tokens due to the non-uniform centroid allocation of \tac, we normalize the residual vectors and store their norms separately. The resulting normalized residuals are more homogeneous and therefore easier to compress effectively.

\subsection{Hierarchical Index for Optimized Multivector Retrieval}
\label{sec:tachiom}

The speedups achieved by \tac over $\kappa$-means allow us to significantly scale up the number of centroids used. This fine-grained clustering significantly improves the approximation provided by centroids alone, thus eliminating the need for full token scoring (centroid + residual) during the gather phase, differently from existing approaches~\cite{emvb,igp,warp}. Furthermore, by decoupling the gather phase from the residual scores calculation, we can specialize the refinement stage with a Product Quantization layout optimized specifically for document-level $MaxSim$ evaluation.
Building on these observations, we introduce \tachiom\ (Token-Aware Clustering and Hierarchical Indexing for Optimized Multivector retrieval), a retrieval architecture that combines an \hnsw proximity graph over centroids for efficient gathering with a cache-optimized PQ layout for efficient refinement.
\tachiom operates in two main phases: first, it explores centroids to select a set of candidate documents; in the second phase, it uses both centroids and residuals to compute the full $MaxSim$ scores of the candidate documents and provide the final ranking.

\vspace{1mm}
\noindent \textbf{Index Structure}. 
We build an \hnsw~\cite{hnsw} graph over the centroid set; each centroid $\mathbf{c}_i$ maintains an inverted list $\mathcal{L}_i$ containing the document IDs of all tokens assigned to $\mathbf{c}_i$. This design enables document-level scoring during the gather phase without accessing individual token vectors.

\vspace{1mm}
\noindent \textbf{Gather Phase: Centroid-based Document Scoring}.
For each query token $\mathbf{q}_i$ of a query $q$: 

\begin{enumerate}
    \item we traverse the graph to identify the top-$\kappa_c$ nearest centroids;
    \item for each retrieved centroid $\mathbf{c}_j$, we iterate over its inverted list $\mathcal{L}_j$;
    \item for documents appearing in multiple lists, we retain only the highest centroid similarity: $\tilde{s}_i(d) = \max_{\{j \,:\, d \in \mathcal{L}_j\}} \langle \mathbf{q}_i, \mathbf{c}_j \rangle$;
    \item we accumulate partial scores across all query tokens: $\tilde{S}(q, d) = \sum_{i=1}^{n_q} \tilde{s}_i(d)$.
\end{enumerate}
 
This produces an approximate $MaxSim$ score using only centroid-level interactions. Documents appearing in inverted lists for multiple query tokens naturally accumulate higher scores, capturing the multi-token matching signal without decompressing token vectors. Before the next phase, we truncate the candidate set to the top $\kappa_d$ documents by partial score $\tilde{S}(q,\cdot,)$. We also apply adaptive filtering to further reduce the candidate set, adopting the \emph{Candidates Pruning} (CP) strategy proposed in \cite{martinico2026multivector}, regulated by a parameter $\alpha$.

\begin{figure}[tb!]
\centering
\includegraphics[width=0.475\textwidth]{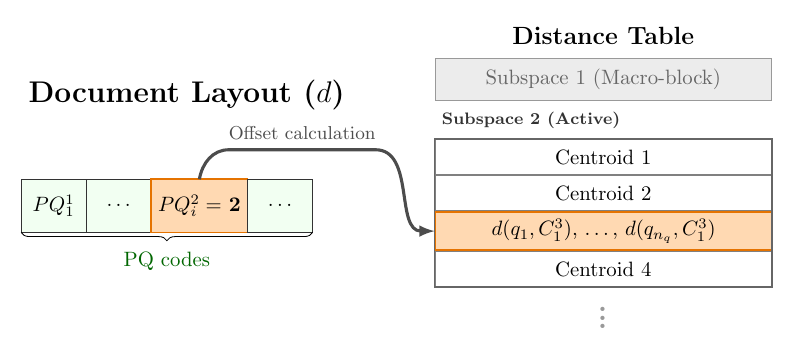}
\vspace{-6mm}
\caption{Distance tables layout with 4 centroids per subspace. Evaluating code $j=2$ for subspace $m=2$, document token $i$.}\label{fig:dist_tables} 
\end{figure}

\vspace{1mm}
\noindent \textbf{Refine Phase: PQ-Optimized $\mathbf{MaxSim}$ Computation}.
For the pruned candidate set, we compute full $MaxSim$ scores using both centroids and PQ-compressed residuals. We adopt specialized data layouts optimized for cache locality and SIMD vectorization.

\textit{Document Layout}: For each document $d$, we store all tokens' centroid IDs contiguously, followed by all PQ codes: $[\mathbf{c}_1^d, \ldots, \mathbf{c}_{n_d}^d \mid \text{PQ}_1^{1}, \ldots, \text{PQ}_{n_d}^{M}]$, where $n_d$ is the number of tokens in document $d$ and $M$ is the number of PQ subspaces. This enables streaming evaluation: we first accumulate all centroid contributions in one pass, then refine with residuals in a second pass.

\textit{Distance Table Layout}: PQ requires precomputing distance tables between query tokens and PQ centroids for each PQ subspace. A naive multivector PQ implementation processes each query token independently, resulting in a separate distance table per token. However, because a document token's quantization is fixed, all $n_q$ query tokens must evaluate their distance against the \emph{same} PQ centroid for any given subspace. Consequently, the standard memory layout necessitates $n_q$ scattered memory accesses across $n_q$ disjoint tables, creating a severe memory-bandwidth bottleneck. To eliminate this overhead and explicitly exploit the shared access pattern inherent to the MaxSim operator, we reorganize the precomputed distances into a cache-optimized three-level hierarchy:
\begin{itemize}
    \item \emph{Macro-blocks}: indexed by PQ subspace ($M$ blocks for $M$ subspaces);
    \item \emph{Centroid blocks}: indexed by PQ centroid ID ($2^b$ blocks for $b$-bit codes);
    \item \emph{Query token micro-blocks}: containing $n_q$ consecutive distances between each query token and the same PQ centroid on that subspace.
\end{itemize}

When evaluating a document token's PQ code $j$ for subspace $m$, our layout enables retrieving the distances to all $n_q$ query tokens as a single contiguous micro-block (Figure~\ref{fig:dist_tables}). This avoids the $n_q$ scattered memory lookups of the standard layout, yielding up to a $3.8\times$ speedup in residual distance computation compared to a standard PQ implementation.

\section{Experiments}
\label{sec:exp}
We perform experiments on \msmarcovo~\cite{msmarco} (8.8M passages, 598M token vectors, 6,980 dev.small queries) for in-domain evaluation and \lotte-pooled~\cite{colbertv2} (2.4M passages, 266M token vectors, 2,931 search/dev queries) for out-of-domain evaluation. We use \colbertvt~\cite{colbertv2} as the encoder with a maximum of 180 tokens per document for \lotte. We evaluate the retrieval results using the standard metrics for each dataset: MRR@10 for \msmarcovo and Success@5 for \lotte. Experiments run on an Intel Xeon Silver 4314 CPU @ 2.40GHz with 64 threads. Our implementation is written in Rust \texttt{1.92.0-nightly} on top of the \kannolo~\cite{kannolo} library. Clustering exploits all 64 threads in parallel, while retrieval experiments are executed sequentially on a single core.

\subsection{Clustering Evaluation}
We first evaluate \tac's efficiency and effectiveness on \msmarcovo against two $\kappa$-means baselines: \faiss~\cite{douze2024faiss}, which is one of the most widely-adopted $\kappa$-means implementations, and \fkmeansrs, an efficient Rust implementation of the \fkmeans library~\cite{fastkmeans2025}. Since our implementation uses \kannolo's AVX2-optimized single-vector distance kernels without production-ready external libraries such as Intel MKL~\cite{IntelMKL}, we evaluate two \faiss\ configurations: generic AVX2 (comparable to our setup) and MKL-enabled.

For \tac, we set tail handling thresholds $\mu=128$ and $\tau=256$, and bounding parameters $\varepsilon=4$ and $\theta=39$ based on preliminary experiments. We compress residuals using PQ with $M=32$ subspaces and 8-bit codes per subspace. The number of $\kappa$-means iterations is set to $10$ for both standard $\kappa$-means and \tac.

Clustering time measurements include (i) centroids computation and (ii) centroid assignments for all vectors in the dataset. The sample size for $\kappa$-means is set to 10M vectors; we assume a 24-hour budget for algorithms' running time. 
We assess clustering quality by measuring retrieval effectiveness when using the compressed representations to rerank candidates. In order to isolate vector approximation from index approximation, we do not rely on \tachiom for retrieval. 
Following the methodology in \cite{martinico2026multivector}, we rerank the top 1,000 documents retrieved by \splade \cocond~\cite{cocondenser} using the RerankIndex with all optimizations disabled---specifically removing early exit and pruning mechanisms---and return the top 10 results for each query. This approach is designed to emulate an exhaustive search while maintaining feasible query times, allowing us to precisely assess our centroids' approximation quality. 

\begin{figure}[tb]
    \centering
    \includegraphics[width=0.95\columnwidth]{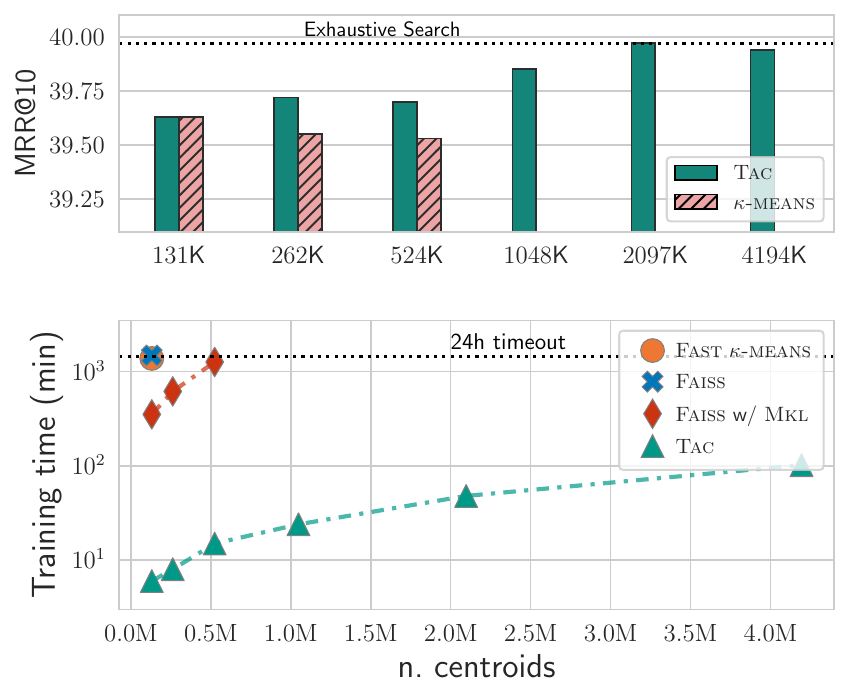}
    \vspace{-4mm}
    \caption{\msmarcovo. Top plot: MRR@10 for 1,000 candidates reranking. Bottom plot: clustering time in minutes.}
    \label{fig:placeholder}
\end{figure}

The results of this evaluation are reported in Figure \ref{fig:placeholder}. As shown in the bottom plot, \tac achieves training times orders of magnitude lower than all competing methods, even when they benefit from Intel MKL acceleration. 
Specifically, \tac yields speedups up to $84\times$ (\faiss with MKL), $247\times$ (\faiss), and $230\times$ (\fkmeansrs) over the competitors.
Our method clusters nearly 600 million 128-dimensional vectors into 262K clusters in just 8 minutes, potentially benefiting prior methods such as \plaid, \emvb, \warp, and \igp, which all rely on this same centroid budget. Furthermore, \tac scales seamlessly to over 4M centroids in only 102 minutes, while the 24-hour timeout prevents the competitors to surpass $131$K centroids (\faiss and \fkmeansrs) and $524$K (\faiss with MKL).

The top plot of Figure \ref{fig:placeholder} compares the MRR@10 achieved by \tac and $\kappa$-means on \msmarcovo across different centroid budgets (both with normalized residuals for a fair comparison). We report the MRR@10 achieved by exhaustive search over \colbertvt as reference\cite{plaid_repro}. This plot highlights that \tac not only offers substantial speedups but also delivers equal or superior approximation quality at a fixed centroid budget, confirming that our allocation strategy effectively improves centroid assignment and validating the benefits of token-aware design for retrieval effectiveness.



\subsection{Retrieval Evaluation}
We evaluate the efficiency-effectiveness trade-off offered by \tachiom by comparing it to three state-of-the-art~\cite{martinico2026multivector} multivector retrieval methods: \emvb~\cite{emvb}, \warp~\cite{warp} and \igp~\cite{igp}.
We report the best (lowest) average query time for \tachiom and all selected competitors at different metrics cut-off points. 

For \tachiom, we set the number of centroids to $2^{21} \approx 2$M for \lotte and $2^{22} \approx 4$M for \msmarcovo. The \hnsw graphs over centroids are built with $M=32$ neighbors and $ef_c=1500$.
At search time, we perform grid search over: number of retrieved centroids per query token $\kappa_c \in \{15, 20, 40, 80, 100, 120\}$, maximum candidate documents $\kappa_d \in \{250, 500, 1000, 2000, 4000\}$, and candidate pruning threshold $\alpha \in \{0.35, 0.4, 0.45, 0.5\}$, with \hnsw search parameter $ef_s = \frac{3}{2} \kappa_c$. For competitors, we follow the grid search procedures from~\cite{martinico2026multivector}.
All the methods use the same residuals size, i.e. $32$ bytes per token vector. This is achieved with PQ32 or its variants for \tachiom and \emvb, and with $2$ bits per component for the Scalar Quantization employed by \warp and \igp.

\begin{table}[tb]
\centering
\caption{Average query time (ms). ``$-$'' indicates the technique does not reach the effectiveness level.}
\label{tab:retrieval_comparison}
\vspace{-3mm}
\small
\setlength{\tabcolsep}{9pt} 
\begin{tabular}{@{} l *{3}{c} @{\hspace{3em}} *{2}{c} c @{\hspace{2em}}} 
\toprule
& \multicolumn{3}{c}{\textbf{\msmarcovo} (MRR@10)} & \multicolumn{3}{c}{\textbf{\lotte} (Success@5)}\\
\cmidrule(lr){2-4} \cmidrule(lr){5-7}
& 39.0 & 39.3 & 39.6 & 67.5 & 68.0 & 68.5 \\
\midrule
\warp    & 98 \rlap{\hspace{1pt}\scriptsize\color{gray}9.8$\times$} & 98 \rlap{\hspace{1pt}\scriptsize\color{gray}7.0$\times$} & -- & 49 \rlap{\hspace{1pt}\scriptsize\color{gray}4.4$\times$} & -- & -- \\
\igp     & 72 \rlap{\hspace{1pt}\scriptsize\color{gray}7.2$\times$} & -- & -- & 48 \rlap{\hspace{1pt}\scriptsize\color{gray}4.3$\times$} & -- & -- \\
\emvb    & 55 \rlap{\hspace{1pt}\scriptsize\color{gray}5.5$\times$} & \phantom{1}56 \rlap{\hspace{1pt}\scriptsize\color{gray}4.0$\times$} & 57 \rlap{\hspace{1pt}\scriptsize\color{gray}3.8$\times$} & 54 \rlap{\hspace{1pt}\scriptsize\color{gray}4.9$\times$} & 55 \rlap{\hspace{1pt}\scriptsize\color{gray}3.9$\times$} & 59 \rlap{\hspace{1pt}\scriptsize\color{gray}2.5$\times$} \\
\textbf{\tachiom} & \textbf{10} \phantom{\rlap{\hspace{1pt}\scriptsize\color{gray}2.2$\times$}} & \phantom{1}\textbf{14} \phantom{\rlap{\hspace{1pt}\scriptsize\color{gray}2.2$\times$}} & \textbf{15} \phantom{\rlap{\hspace{1pt}\scriptsize\color{gray}2.2$\times$}} & \textbf{11} \phantom{\rlap{\hspace{1pt}\scriptsize\color{gray}2.2$\times$}} & \textbf{14} \phantom{\rlap{\hspace{1pt}\scriptsize\color{gray}2.2$\times$}} & \textbf{23} \phantom{\rlap{\hspace{1pt}\scriptsize\color{gray}2.2$\times$}} \\
\bottomrule
\end{tabular}
\end{table}

Our results are reported in Table \ref{tab:retrieval_comparison}. \tachiom exhibits superior efficiency across all metric cutoffs, achieving speedups ranging from $2.5\times$ to $9.8\times$ compared to the baselines. \tachiom is up to $5.5\times$ faster than the best available data structure \emvb and 
matches its peak effectiveness despite employing standard Product Quantization (PQ). \emvb, conversely, relies on computationally expensive PQ variants, namely JMPQ~\cite{jmpq}---a supervised method that jointly trains centroids, PQ, and the query encoder---on \msmarcovo and OPQ~\cite{opq} on \lotte. 
These results demonstrate that \tachiom, through its careful centroid selection and ability to scale centroids via \tac, rivals the effectiveness of both JMPQ and OPQ without incurring their additional training overhead.

\section{Conclusions and Future Work}
\label{sec:conclusion}
In this work, we presented \tac, a clustering algorithm that decomposes the global clustering problem into independent per-token subproblems and allocates centroids to tokens based on their frequency and semantic variance, achieving up to $247\times$ speedup over $k$-means and improving effectiveness. 
On top of this, we developed \tachiom, a retrieval architecture that leverages high-granularity centroids to efficiently gather candidates, bypassing expensive token-level interactions. \tachiom achieves a $9.8\times$ speedup over state-of-the-art methods. 
Together, these contributions unlock the practical utility of multivector retrieval for large-scale datasets.
Future work will assess the generalizability of \tac\ beyond \colbertvt\ and evaluate \tachiom\ on datasets substantially larger than \msmarcovo.
Additionally, by leveraging the superior approximation quality of \tac centroids, we intend to investigate higher residual compression ratios.

\section*{Acknowledgements}
This research was conducted as part of the ENRESFOOD project, funded under the FutureFoodS partnership. The project is co-funded by the Austrian Science Fund (FWF; 10.55776/KIN4661525), The Dutch Ministry of Agriculture, Fisheries, Food Security and Nature (BO-43-219-026), the Italian Ministry of University and Research (MUR), the German Federal Ministry of Research, Technology and Space (031B1717), and the European Union’s Horizon Europe research and innovation programme via FutureFoodS, the European Partnership for a Sustainable Future of Food Systems, co-funded under Grant Agreement No. 101136361.

\bibliographystyle{ACM-Reference-Format}
\bibliography{sample-base.bib}

\end{document}